\documentclass[12pt]{article}

\usepackage[utf8]{inputenc}
\usepackage{textcomp}
\usepackage{comment}
\usepackage{natbib}
\usepackage{amsmath}
\usepackage{amsfonts}
\usepackage{amssymb}
\usepackage{mathtools}
\usepackage[a4paper, margin=1 in]{geometry}
\usepackage{multirow}
\usepackage{multicol}
\usepackage{graphicx}
\usepackage{float}
\usepackage{enumitem}
\usepackage{ntheorem}
\usepackage{MnSymbol}
\usepackage{theoremref}
\usepackage{pifont}
\usepackage{fullpage}
\usepackage{tikz}
\usepackage{physics}
\usepackage{type1cm}
\usepackage{wrapfig}
\usepackage[font=footnotesize,labelfont=bf]{caption}
\usepackage[ruled,vlined]{algorithm2e}
\usepackage{listings}
\usepackage{xcolor}
\usepackage{bbm}
\usepackage{pdfpages}
\usepackage{CJK}
\usepackage{longtable}
\usepackage{titlesec}
\usepackage{subcaption}
\usepackage{hyperref}
\usepackage{csquotes}

\usepackage{theorem}
\newcounter{theorem}

\newtheorem{result}{Result}
\usepackage[figuresright]{rotating}
\usepackage{comment}
\usepackage{multirow, multicol}
\usepackage{natbib}
\usepackage{hyperref}
\usepackage{mathtools}

\renewcommand{\baselinestretch}{1.65} 

\newcommand{\customcite}[1]{\citeauthor{#1}, \href{cite.#1}{\textcolor{blue}{\citeyear{#1}}}}

\newcommand{\customcitetwo}[1]{\citeauthor{#1} (\href{cite.#1}{\textcolor{blue}{\citeyear{#1}}})}

\newcommand{\customcitet}[2]{(\href{cite.#1}{\citeauthor{#1}}, \href{cite.#1}{\textcolor{blue} {\citeyear{#1}}}; \href{cite.#2}{\citeauthor{#2}}, \href{cite.#2}{\textcolor{blue} {\citeyear{#2}}})}
\newcommand{\customcitefr}[4]{(\href{cite.#1}{\citeauthor{#1}}, \href{cite.#1}{\textcolor{blue} {\citeyear{#1}}}; \href{cite.#2}{\citeauthor{#2}}, \href{cite.#2}{\textcolor{blue} {\citeyear{#2}}}; \href{cite.#3}{\citeauthor{#3}}, \href{cite.#3}{\textcolor{blue} {\citeyear{#3}}}; \href{cite.#4}{\citeauthor{#4}}, \href{cite.#4}{\textcolor{blue} {\citeyear{#4}}})}

\newcommand{\customref}[1]{\textcolor{blue}{\ref{#1}}}

\newcommand{\estimates}{\mathrel{\widehat{=}}}

\title{\bf \Large Dir-SPGLM: A Bayesian semiparametric GLM with data-driven reference distribution
}

\author{Entejar Alam$^{1}$,  Peter Müller$^{1,2}$, and Paul J. Rathouz$^{3,1}$\\
$^{1}$Department of Statistics and Data Sciences, \\
$^{2}$Department of Mathematics, \\
$^{3}$Department of Population Health, \\
The University of Texas at Austin, TX, USA}

  \date{}

\begin{document}

\def\spacingset#1{\renewcommand{\baselinestretch}%
{#1}\small\normalsize} \spacingset{1}
  \maketitle
  \bigskip
  \begin{abstract}
  \noindent The recently developed semi-parametric generalized linear model (SPGLM) offers more flexibility as compared to the classical GLM by including the baseline or reference distribution of the response as an additional parameter in the model. However, some inference summaries are not easily generated under existing maximum-likelihood based inference (ML-SPGLM). This includes uncertainty in estimation for model-derived functionals such as exceedance probabilities. The latter are critical in a clinical diagnostic or decision-making setting. In this article, by placing a Dirichlet prior on the baseline distribution, we propose a Bayesian model-based approach for inference to address these important gaps. We establish consistency and asymptotic normality results for the implied canonical parameter. Simulation studies and an illustration with data from an aging research study confirm that the proposed method performs comparably or better in comparison with ML-SPGLM. The proposed Bayesian framework is most attractive for inference with small sample training data or in sparse-data scenarios.
\end{abstract}
\noindent%

\def\spacingset#1{\renewcommand{\baselinestretch}%
{#1}\small\normalsize} \spacingset{1.65}
\section{INTRODUCTION \label{section intro}}

We introduce a Bayesian model-based approach to a recently developed semi-parametric extension of the classical generalized linear model (GLM; \customcite{mccullagh1989generalized}). \customcitetwo{rathouz2009generalized} and \customcitetwo{wurm2018semiparametric} proposed a semi-parametric generalized linear model (SPGLM) for regression of a scored ordinal outcome or other discrete distributions with finite support, for example, a count of symptoms or of items endorsed.  In those earlier works, the 
SPGLM model is referred to as generalized linear density ratio model.   The main novelty of the proposed framework is that the model-based approach facilitates full predictive inference, including clinically important exceedance probabilities and inference for design questions. We are introducing Bayesian Dirichlet ordinal regression as a novel statistical data analysis technique.

The defining characteristic of the SPGLM is a nonparametric baseline or reference distribution $f_0$ that replaces a parametric specification in the classical GLM such as binomial or Poisson distribution. Keeping $f_0$ nonparametric instead, the analyst needs to only specify the linear predictor and link function, even avoiding a variance function. Model specification is less onerous than even with quasilikehood (QL) models, while still yielding a valid likelihood function. Beyond the initial proposal by \customcitetwo{rathouz2009generalized}, which focused primarily on the finite support case, 
 \customcitetwo{huang2014joint}
 went on to mathematically characterize the SPGLM in the infinite support case, and \customcitetwo{maronge2023generalized} have shown how the model can be used to handle outcome-dependent or generalized case-control sampling.  

Despite these developments, there are still many important gaps in the literature on inference procedures that have been developed for these models. These include inference for application-driven functionals of the fitted models such as exceedance probabilities, which are crucial in clinical diagnostic (\customcite{paul2021dynamics}), natural hazard detection \customcitet{kossin2020global}{tajima2021empirical}, financial risk management (\customcite{taylor2016using}) or in general, any decision-making setting. For these quantities, approach based on maximum likelihood estimation cannot easily generate inference, as well as a full description of all relevant uncertainties, as would be needed for a design problem.  These comments are elaborated below using a motivating case study in Section \customref{sec2}.  The main aim of this paper is to develop a Bayesian approach to inference under the SPGLM. We will refer to the proposed Bayesian extension of the model as the Dir-SPGLM.



Under a Bayesian approach, the baseline distribution $f_0$ of the GLM becomes one of the unknown parameters. A complete Bayesian inference model therefore requires a prior probability model $p(f_0)$ on $f_0.$ Prior models on random probability measures are known as nonparametric Bayesian models (BNP) (\customcite{ghosal2017fundamentals});
the most widely used one being the Dirichlet process (DP) prior \customcitet{ferguson1973bayesian}{ghosal_2010}.
One characterization of the DP prior is as a limit of symmetric Dirichlet priors on finite sample spaces (\customcite{GreenRichardson01}).
In fact, the DP prior reduces to a Dirichlet distribution in the case of a finite sample space, as is the case in the SPGLM development here. 

Many BNP models in the recent literature introduce variations of the DP prior to dependent DP (DDP) models for families of dependent random probability measures \customcitet{maceachern;2000}{quintana2022dependent}. The dependency here is introduced through a set of predictors $x$; exploring this dependence structure is of substantial interest in biomedical and public health studies. In particular, DDP family include models under which the dependent random probability measures share the same atoms but dependent weights through $x$ (common atoms DDP) and others that share the same weights but dependent atoms through $x$ (common weights DDP). The proposed Dir-SPGLM is a special case of a common atoms DDP prior for a finite sample space.

The rest of the article is organised as follows. Section \customref{sec2} motivates the problem, and provides a detailed data description for the ``Assets and HEAlth Dynamics among the oldest old" (AHEAD) study (\customcite{soldo1997asset}), which is later used in an illustrative analysis. In Section \customref{sec3}, we fully specify and develop our Dir-SPGLM model followed by theoretical results, and a discussion on model properties and induced prior distributions. Section \customref{sec4} outlines a Markov chain Monte Carlo (MCMC) algorithm for posterior simulation within the Dir-SPGLM framework. In Section \customref{sec5}, we illustrate our method on various simulated datasets and examine frequentist operating characteristics using replicated simulation.  Section \customref{sec6} reports inference for the AHEAD study dataset,  covering performance on both small and large samples. In Section \customref{sec7}, we conclude with a discussion and future directions. Proofs and additional data illustrations are included in the Supplementary Information.

\section{THE AHEAD STUDY}\label{sec2}
The AHEAD study (\customcite{soldo1997asset}) collected nationwide longitudinal data from an initial sample of over $7,000$ participants aged $70$ years and older at baseline, including their spouses and household partners. This study's primary goals involved tracking the transitions in physical, functional, and cognitive health and investigating how changes in late-life physical and cognitive health relate to patterns of dissaving and income flow. For the analysis here, we focus specifically on the baseline data collected in 1993. The complete data sample size is $n = 6,441$. 

\begin{figure}[ht]
    \centering
    \includegraphics[height = 7cm, width = 11cm]{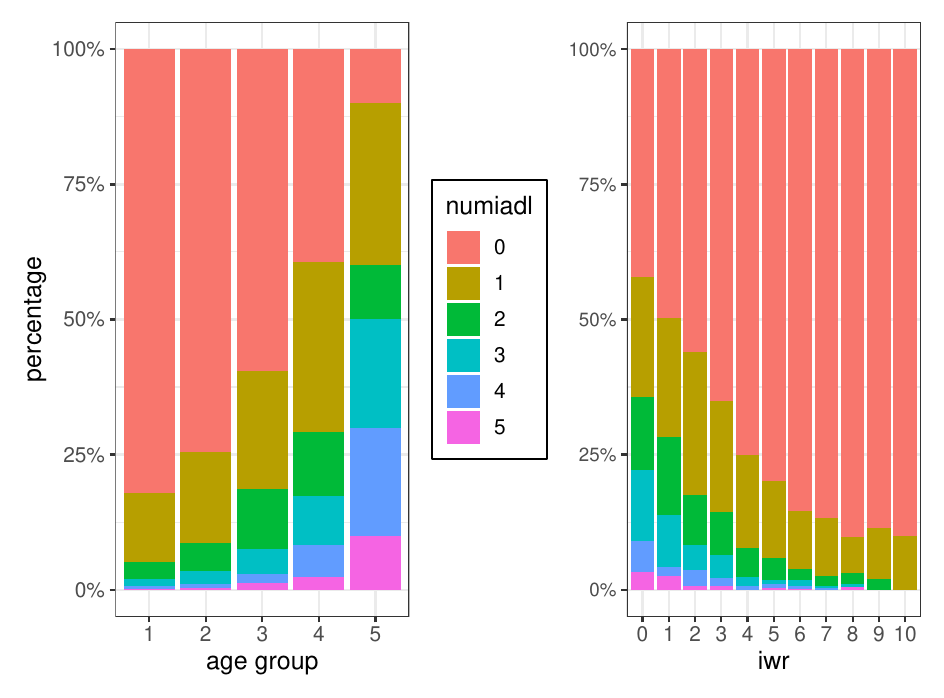}
    \caption{Variables \textit{numiadl} (difficulty count of daily activities), \textit{iwr} (immediate word recall), and \textit{age} from the AHEAD data set; age group (years): 1 (70-75), 2 (76-82), 3 (83-89), 4(90-96), and 5 (96+ ).\label{fig1}}
\end{figure}

Figure \customref{fig1}, we present the data for the variables \textit{numiadl} (difficulty count of daily activities), \textit{iwr} (immediate word recall), and \textit{age}. For this graphic, we have coded \textit{age} into $5$ ordinal levels. We observe evidence for a positive association between the variables \textit{numiadl} and \textit{age}, while \textit{numiadl} shows a negative association with \textit{iwr}. Hence, modeling the scored ordinal variable \textit{numiadl} based on \textit{age}, \textit{iwr} and other variables is of particular interest.
Some options for modeling this response, with their limitations, include:  A Poisson model, the problem here being that the distribution is limited to only 0 through 5, and has considerable overdispersion at 0.  Similarly, one could use a binomial distribution; this model would also suffer from overdispersion.  Of course, a quasilikelihood approach could be taken, but this would not yield full distributional inferences (such as exceedance probabilities) because the QL model is essentially a model for the mean.  Finally, ordinal data regression could be used; our concern with this is that the regression parameters (in for, e.g., ordinal logistic regression) are for log-cumulative odds ratios, which can be challenging to interpret to applied audiences; our model provides regression parameters for the mean response, interpreted in terms of the analyst-chosen link function.

\customcitetwo{rathouz2009generalized} analysed the same data using maximum likelihood in the SPGLM, as stated in (\customref{eq1})--(\customref{eq3}) below, yielding valid inferences for regression parameters $\beta$. It has been challenging, however, to extract both estimates and uncertainty quantifications for either the centering distribution or for model-derived quantities such as exceedance probabilities $p(y \geq y_0\mid x)$, which depend jointly on $\beta$ and the baseline density. The main reason for this limitation is that the joint sampling variance-covariance matrix would be needed for all model parameters, a quantity that has been difficult to extract. Even if available, manipulating it via, e.g., the ``delta method" may prove challenging, and may fail to provide sufficiently good approximation to valid inferences for reasonable sample sizes. An alternative to addressing these critical gaps is a Bayesian approach, which always provides full uncertainty quantification for any model-derived quantity or event of interest via the joint posterior sample of all the model parameters.

The proposed semiparametric Bayesian approach implements this goal and prepares the model for future extensions to multivariate or repeated measurement models by introducing appropriate random effects and other latent variables. The concept behind the proposed Dir-SPGLM is straightforward. Existing semi-parametric frequentist methods (\customcite{rathouz2009generalized}; \customcite{wurm2018semiparametric}) lack a Bayesian perspective, while parametric Bayesian approaches (\customcite{gelfand1999identifiability}) are restricted by their parametric nature. To overcome the limitations of both, and simultaneously employ their respective strengths, we combine these two approaches to develop a novel semi-parametric Bayesian model known as Dir-SPGLM.

\section{MODEL DEVELOPMENT}\label{sec3}

\subsection{The Dir-SPGLM Model}

Consider a GLM  
\begin{align}
  y \sim f(y \mid x) = \exp\left\{\theta y - b(\theta)\right\} f_0(y) \label{eq1} 
\end{align}
with response $y \in \mathcal{Y} \subset \mathcal{R}$ and a p-dimensional covariate vector $x$. Here, $f_0$ is the baseline density and $b(\cdot)$ is the log-normalization constant
\begin{align}
b(\theta) = \log \int_{\mathcal{Y}} \exp (\theta y) f_0(y)  dy \ . \label{eq2}  
 \end{align}
In the classical GLM, the baseline distribution $f_0(y)$ is assumed to be in a parametric family---in the proposed semi-parametric model it becomes an unknown parameter.  As in the classical GLM, $\eta = x^T\beta$ is a linear predictor, $g$ is a link function, and $\mu = E(y\mid x) = g^{-1}(\eta)$. For fixed $f_0,$ the expectation $\mu$ implicitly determines $\theta$ by the equation
\begin{align}
\mu = E(y \mid x) = b^{\prime}(\theta) = \frac{\int_{\mathcal{Y}} y \exp (\theta y) f_0(y)  dy}{\int_{\mathcal{Y}} \exp (\theta y) f_0(y)  dy} \ ,\label{eq3}
\end{align}
where we note that $b^{\prime}(\theta)$  is a strictly increasing function of $\theta.$ The free parameters in the model are $\beta \in \mathcal{R}^p$ and $f_0 \in \mathcal{F}$, with $\mathcal{F}$ defined below. By contrast, $\theta = \theta (\beta, f_0; x)$ is a derived parameter based on (\customref{eq3}). We can write the solution for $\theta$ as a function of $\mu$, denoted as ${b^\prime}^{-1}(\mu; f_0)$, which additionally depends on $f_0$, in addition to depending on $\beta$, implicitly through $\mu = x^T\beta$.

Here, we consider $y$ to be a scored ordinal outcome, by which we mean an ordinal outcome in which the mean response of (given) scores $s_\ell$ is of interest. In this case:
$\mathcal{Y} = \{s_1, \dots, s_k\}$, where $s_\ell < s_{\ell + 1}$, and the integration in (\customref{eq2}) and (\customref{eq3}) becomes a finite sum. In this setting $f_0$ is restricted to $\mathcal{F} := \{f : f(s_\ell) \in [0, 1] \mbox{ and } \sum_{\ell=1}^k f(s_\ell) = 1 \}$, i.e., the $k-1$ dimensional simplex. For the remaining discussion we restrict attention to this case of scored ordinal outcomes. 

\noindent Extending the SPGLM to a Bayesian inference model requires prior distributions on the parameters, $\beta = (\beta_1, \dots, \beta_p)$ and $f_0 \equiv \left\{f_0\left(s_1\right), \ldots, f_0\left(s_k \right) \right\}$.
We use
\begin{align}
  \beta \sim N_p (0, \mathcal{I}_p), \mbox{ and }
  f_0 \sim Dir(\alpha H) \equiv Dir \left(\alpha H(s_1), \dots, \alpha H(s_k) \right), \label{eq4}  
\end{align}
where $\alpha > 0$ is a user-specified concentration parameter, and $H(s_\ell)$ is the probability mass at $s_\ell$ under a user-specified centering distribution $H$, implying $E(f_0)=(H)$. The Dirichlet prior on $f_0$ can equivalently be represented as $f_0(\cdot) = \sum_{\ell=1}^k u_\ell \delta_{s_\ell}(\cdot)$, where $u_\ell \propto \widetilde{u}_\ell$ (i.e. 
$u_\ell = \widetilde{u}_\ell \big/ \sum_{\ell^\prime =1}^k \widetilde{u}_{\ell^\prime}$)
with $\widetilde{u}_\ell \sim Gamma\left(\alpha H(s_\ell), \phi \right),$ independently across $\ell$, for an arbitrary scale parameter $\phi$.
This implies the following model for $f(y \mid x)$ in (\customref{eq1}) 
\begin{align}
    f(y \mid x)  = 
\exp\left\{\theta y - b(\theta) \right\} \sum_{\ell=1}^k u_\ell \delta_{s_\ell} (y) = \sum_{\ell=1}^k \left[\exp\left\{\theta s_\ell -b(\theta)\right\} u_\ell\right] \delta_{s_\ell} (y) = \sum_{\ell=1}^k  w_{x\ell} \delta_{s_\ell} (y), \label{eq5}
\end{align}
where $w_{x \ell} = \exp\left\{\theta s_\ell - b(\theta)\right\}  u_\ell$ varies with $x$ (recall that $\theta$ is a function of $\mu = x^\prime\beta$ and $f_0$). That is, $w_{x\ell}$ is also a function of $\beta$ and $f_0$. Let, $w_x$ denote the vector of weights, $w_x = \left(w_{x1}, \ldots, w_{xk}\right)$ with $\sum_{\ell=1}^k w_{x\ell} = 1$. We refer to the distribution of $w_x$ as a tilted Dirichlet distribution.  In general, a tilted Dirichlet random vector
$v = (v_\ell;\; \ell=1,\ldots,k)$ is a modified Dirichlet random vector $\widetilde{v} \sim Dir(c)$ with hyperparameter $c = (c_\ell;\; \ell=1,\ldots,k)$ that
is exponentially tilted as $v_\ell \propto e^{\theta s_\ell} \widetilde{v}_\ell$ with
respect to given scores $s_\ell \in s = (s_{\ell^\prime};\; \ell^\prime =1,\ldots,k)$ to fix a mean $\mu = \sum_\ell
s_\ell v_\ell \big/ k$ (with respect to the same scores).
That is, we use a Dirichlet distribution to generate random weights
for a scored ordinal random variable $y$ with probability function
$p(y=s_\ell) = v_\ell$, subject to a given mean $\mu$ (of $y$). The
desired mean is achieved by exponential tilting of a Dirichlet random vector. We write $v \sim tDir(c, \mu, s)$.

Finally, we comment on (likelihood) identifiability under the proposed model. Equation (\customref{eq4}), without additional restrictions, would suffer from a lack of likelihood identifiability. A given pair ($f_0, \theta_0$) could always be replaced by another arbitrary pair $(f_1, \theta_1)$ with $f_1 \propto f_0 \cdot \exp(c y)$ and  $\theta_1 = (\theta_0-c)$, leaving the joint likelihood invariant. \customcitetwo{rathouz2009generalized} introduce a constraint on $f_0$ as $E_{f_0}(y) = \sum_{\ell = 1}^k s_\ell f_0(s_\ell) = \mu_0$ for a user-specified $\mu_0.$ Let then $f^\star_0$ denote the $f_1$ in that set of equivalent models with $E_{f^\star_0}(y)=\mu_0$. Let $\mathcal{F}^\star = \{f \in \mathcal{F} : E_f(y) = \mu_0\}$ denote the restricted set of distributions on $\mathcal{Y}$ with the prescribed expectation $\mu_0$. A (prior) probability measure on $\mathcal{F}$ implicitly defines a prior probability model on the restricted space $\mathcal{F}^\star$. In practice, the choice of $\mu_0$ can be based on prior judgement, or alternatively one can use any measure of central tendency of $y$, such as the mean or median. In addition, choose the hyperparameter $H$ to favor $f_0$ centered around the selected $\mu_0$, in the following sense. Select $H$ to be equal or close to the observed frequency distribution of $y$ such that the mean of $H$ is $\mu_0.$  Then, $E_H(y)=E_H\{ E_{f_0}(y\mid f_0)\}.$ 

In summary, equation (\customref{eq5}) together with the prior model (\customref{eq4}) and the restriction on $f_0$ define an ordinal regression in the form of a family of tilted Dirichlet distributions for $w_x$. We refer to this Bayesian ordinal regression model as Dir-SPGLM. The main addition to the SPGLM (\customcite{rathouz2009generalized})  is the Dirichlet prior on $f_0$ to define the tilted Dirichlet regression (Dir-SPGLM).

\subsection{Model Properties and Implied Prior for $\theta$}

The proposed Dir-SPGLM model has the following property. The proof can be found in Appendix A of the Supplementary Information.
\begin{result}
The exponentially tilted Dirichlet distribution, $tDir(c, \mu, s)$ with $c = \{\alpha H(s_\ell); \ell = 1, \ldots, k\}, s = \{s_\ell; \ \ell = 1, \ldots, k\}$, is outside the Dirichlet parametric family; however, it remains a valid probability distribution defined over the $(k-1)$ dimensional simplex. In the context of (\customref{eq1}), exponential tilting transforms $f_0(s_\ell)$ to $f(s_\ell \mid x) \propto \exp(\theta s_\ell) f_0(s_\ell)$.
\label{tilted-Dir}
\end{result}

\begin{figure}[ht]
    \centering
    \includegraphics[height = 7cm, width = 10cm]{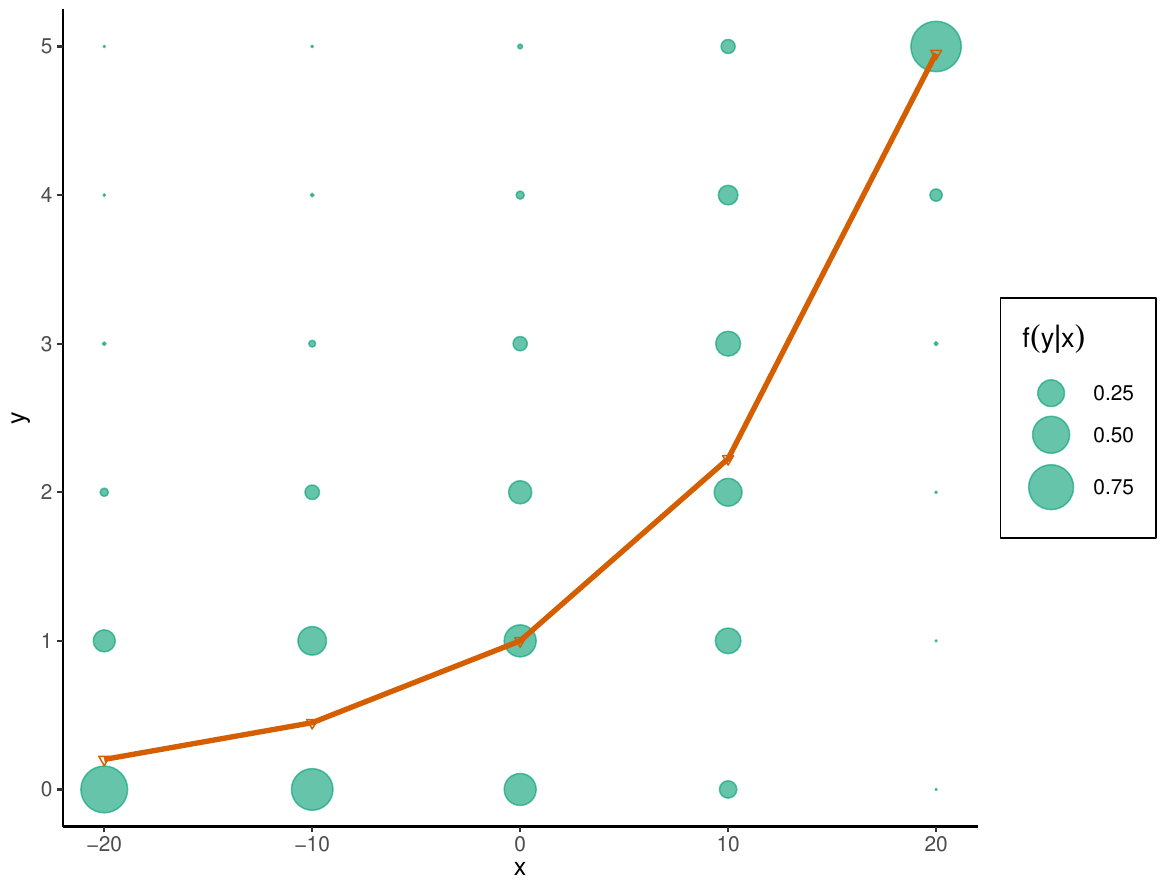}
\caption{Ordinal regression model (\customref{eq1})--(\customref{eq3}). $E(y \mid x)$ is plotted against $x$ (orange line) together with $f(y \mid x)$ (green bullets), where the size of each bullet is proportional to $f(y \mid x)$.\label{fig2}}
\end{figure}

Figure \customref{fig2} illustrates the ordinal regression model (\customref{eq1})$-$(\customref{eq3}), specifying $g$ as the log link and $\mu_0 = 1$. The figure plots $E(y \mid x)$ against $x$ (orange line) together with $f(y \mid x)$ (green bullets). The increasing conditional expectation of the regression mean curve requires $f(y \mid x)$ to become increasingly left-skewed with increasing $x$. For $x = 0$, $E(y \mid x)$ becomes exactly equal to $1$, which results in (\customref{eq1}) representing an i.i.d. sample from $f_0$. However, when $x$ moves away from $0$ in either direction, the distribution of $y$ departs from $f_0$, and becomes more and more left- or right-skewed (relative to $f_0$) due to the tilting factor. Thus, the tilting parameter provides a flexible way to introduce and control skewness (relative to $f_0$) in the Dirichlet distribution, and introduce the desired regression on $x$. 

\noindent For larger $|x|$, the prior distribution of the derived parameter $\theta$ tends to become comparatively flatter (less informative). Conversely, for $x$ closer to $0$, the prior distribution appears to be concentrated around $0$ (more informative). In particular, for $x = 0$ the prior distribution on $\theta$ is degenerate at $0$. The following Result \customref{eqn:theta-prior} characterizes the implied prior on $\theta$ for $x$ close to $0$. The proof can be found in Appendix A of the Supplementary Information.

\begin{result}
\label{eqn:theta-prior}
Under the Dir-SPGLM with prior model (\customref{eq4}), with the additional constraint that $\mu$ is bounded away from the two endpoints $s_1$ and $s_k$, and $f_0 \in \mathcal{F}^\star = \{f^\star_0 \in \mathcal{F} : E_{f^\star_0}(y) = \mu_0\}$, if $g$ and $\mu_0$ are chosen such that as $||x||_2 \to 0$, \ $\mu = g^{-1}(x^T\beta) \stackrel{P}{\rightarrow} \mu_0$, then the derived parameter $\theta = \theta(x; \beta, f_0)$ has the following properties
\begin{enumerate}
    \item[(a)] $\theta \stackrel{P}{\rightarrow} 0$, as $||x||_2 \to 0$, and 
    \item[(b)] $\theta$ is asymptotically normal with mean zero, as $||x||_2 \to 0$.
\end{enumerate}
\end{result}

\section{POSTERIOR SAMPLING AND INFERENCE}\label{sec4}

We implement posterior inference using Markov chain Monte Carlo (MCMC) posterior simulation with Metropolis-Hastings (MH) transition probabilities.

\noindent Let $\{x_i, y_i\}_{i=1}^n$ denote the observed data. For short, we write $f_\ell$ for $f_0(s_\ell)$. Given currently imputed parameters $(\beta, f_1, \dots, f_k) = (\beta^{(t)}, f^{(t)}_1, \dots, f^{(t)}_k)$, MH transition probabilities are defined using the following proposals $\widetilde{\beta}$ and $\widetilde{f}_0$. 

\noindent For $\widetilde{\beta}$, we define a random walk proposal using the Fisher information matrix,
\begin{align}
\widetilde{\beta} \sim \mathcal{N}_p\left(\beta, \ \rho \mathcal{I}^{-1}\left(\beta ; f_1, \dots, f_k\right)\right) 1_A\left(\widetilde{\beta}\right)  \ ,\label{eqn:beta_proposal}
\end{align}
where $A = \left\{\beta \in \mathcal{R}^p:  g^{-1}\left(x_i^T \beta \right) \in [s_1, s_k]\right \}$, $\rho \in (0,1]$ is a tuning parameter, and $\mathcal{I}^{-1}(\beta ; f_1, \dots, f_k)$ is the inverse Fisher information matrix associated with $\beta$. That is,
\[\mathcal{I}\left(\beta ; f_1, \dots, f_k\right) = \sum_{i=1}^n  \frac{x_i x_i^T}{\left(g^{\prime}\left(\mu_i\right)\right)^2 b^{\prime \prime}\left(\theta_i\right)} \ ,\]
$g^{\prime}\left(\mu_i\right) = g^{\prime}\left(g^{-1}\left(x_i^T \beta\right)\right)$, and $b^{\prime \prime}\left(\theta_i\right) = \text{Var}(y_i\mid x_i)$ is given by
\[b^{\prime \prime}\left(\theta_i\right) = \frac{\sum_{\ell = 1}^k s^2_\ell \exp (\theta_i s_\ell) f_\ell}{\sum_{\ell = 1}^k \exp (\theta_i s_\ell) f_\ell} - \left[\frac{\sum_{\ell = 1}^k s_\ell \exp (\theta_i s_\ell) f_\ell}{\sum_{\ell = 1}^k \exp (\theta_i s_\ell) f_\ell} \right]^2\ .\]

\noindent For $\widetilde{f}_0 = (\widetilde{f}_1, \dots, \widetilde{f}_k)$ we use an independence chain proposal based on a weighted empirical distribution:
\begin{align}
(\widetilde{f}_1, \dots, \widetilde{f}_k) \sim  Dir \left(\alpha H(s_1) +
\sum_{i=1}^n \omega_i 1_{\{y_i = s_1\}}, \dots, \alpha H(s_k) + \sum_{i=1}^n \omega_i 1_{\{y_i = s_k\}}\right)  \label{eqn:f0_proposal}
\end{align}
Here, $\omega_i = \frac{c_i}{\sum_{i=1}^n c_i}$, 
where $c_i = \left[\exp\left\{\theta_i y_i - b(\theta_i) \right\}\right]^{-1}$, 
and the value of $\theta_i = \theta_i(\beta, f_1, \dots, f_k; x_i)$ is determined as in (\customref{eq3}) through numerical solution of the equation,
\[\mu_i = g^{-1}\left(x_i^T \beta\right) = \frac{\sum_{\ell = 1}^k s_\ell \exp (\theta_i s_\ell) f_\ell}{\sum_{\ell = 1}^k \exp (\theta_i s_\ell) f_\ell} \ .\]
The motivation for incorporating the weights $\{\omega_i\}_{i = 1}^n$ in the parameters of the above Dirichlet proposal is as follows: If $y_i$ were an i.i.d.\  sample from $f_0$, then (\customref{eqn:f0_proposal}) with $\omega_i = 1$ would define the conditional posterior. The weights are adjusted for the lack of i.i.d.\ sampling (i.e., sampling is condition on $x$) by inverse probability weighting with the additional tilting factor.

We use the two proposals $\widetilde{\beta}$ and $\widetilde{f}_0$ to define the following MH transition probabilities. For $\beta$ we proceed either jointly or one at a time based on the proposals (\customref{eqn:beta_proposal}). The tuning parameter $\rho$ in (\customref{eqn:beta_proposal}) can be adjusted to achieve a well mixing MCMC chain. Also, we sequentially update $f_j$ one at a time keeping others fixed, using the proposals implied by (\customref{eqn:f0_proposal}). 

\section{SIMULATION STUDY}\label{sec5}

We validate the proposed method, and compare its operating characteristics to  ML-SPGLM \customcitet{rathouz2009generalized}{wurm2018semiparametric}. For this we conducted a variety of simulation experiments in which we quantify the degree of (un)biasedness in parameter estimates, coverage probabilities, and statistical efficiency in both small and large sample regimes. Furthermore, we focus on performance in  estimating exceedance probabilities. These often being of special interest in  clinical diagnostic or decision-making settings.

\subsection{Data generation} 
\label{section:5.1}
For the scored ordinal response $y_i$, we take $\mathcal{Y} = \{0, 1, \dots, 5\}$ as the support. The covariate vector is considered as $x_i = (x_{1i}, x_{2i}) \in \mathcal{R}^2$, where $x_{1i} = 1$ and $x_{2i} \stackrel{iid}{\sim} \mathcal{N}(0, 1)$. We then generate $\{y_i\}_{i=1}^n$ conditional on $\{x_i\}_{i=1}^n$ using regression model (\customref{eq1})--(\customref{eq3}), with mean specified as: $\log \mu_i = \log E(y_i \mid x_i) = \beta_0 + \beta_1 x_{2i}$. The simulation truth for the regression parameter is $\beta = (\beta_0, \beta_1) = (-0.7, 0.2)$. For the centering distribution parameter $f_0$, we use these two choices:
\begin{enumerate}
\item [(a)] \underline{Scenario 1:} Truncated Poisson with a right truncation at 5, i.e., $f_0(s) = p_0(s)\big/\sum_{\ell = 0}^5 p_0(\ell),$ $ s \in \mathcal{Y}$. Here $p_0$ is the pmf of a Poisson distribution with mean $\lambda = 1$.
\item[(b)] \underline{Scenario 2:} Zero-inflated truncated Poisson, i.e., $f_0(0) = 3p_0(0)\big/[3p_0(0) + \sum_{\ell = 1}^5 p_0(\ell)]$,
and $f_0(s) = p_0(s)\big/[3p_0(0) + \sum_{\ell = 1}^5 p_0(\ell)], s = 1, \dots, 5$.
\end{enumerate}

These two scenarios are designed to reflect common settings for data with health and behavioral outcomes, namely bounded count or count-like responses, and inflation at one end of the scale. We use $\mathcal{D}_n$ to refer the observed data $\{x_i, y_i\}_{i=1}^n$. For a small sample setting, we consider $n = 25$, and for a large sample setting, $n = 250$. Repeatedly simulating data under this design generates $R$ replicated data sets. We consider $R = 2,\! 000$.  

\subsection{Competing methods}
\label{section:5.2}
We fit both competing inference models for each simulated data set as follows. Considering the prior distributions as in (\customref{eq4}) with $\alpha = 1$ and $H$ as the empirical distribution of $y$, we fit the proposed Dir-SPGLM model by MCMC posterior simulation for $\beta$ and $f_0$ using the transition probabilities described in (\customref{eqn:beta_proposal}) and (\customref{eqn:f0_proposal}). We use tuning parameter $\rho = 1$, and a total of $5,\!000$ MCMC samples are generated for each replicate. We discard the first $2,\!000$ iterations as initial burn-in and use the remaining $B = 3,\!000$ Monte Carlo samples for the following results. We fit the ML-SPGLM model using maximum likelihood (ML) estimation \customcitet{rathouz2009generalized}{wurm2018semiparametric}.

We use the following measures for a comprehensive comparison of the two inference methods. For quantifying frequentist bias (or lack thereof): \textit{Est$_a$} and \textit{Est$_m$}, which respectively are average and median of the parameter estimates across the replicates. For this comparison, we tilt both the posterior samples and ML estimate of $f_0$ to have a common reference mean $\mu_0$. We use $\mu_0$ equal to the simulation ground truth (Section \customref{section:5.1}). We report posterior means as estimates under Dir-SPGLM and maximum likelihood estimates for ML-SPGLM. For statistical efficiency in frequentist terms, we report \textit{RRMSE} (relative root mean square error) and \textit{RL} (relative length) for uncertainty intervals. Both are relative to ML-SPGLM. We report \textit{RRMSE$_m$}, defined as the ratio of median (across replicates) RMSE for Dir-SPGLM versus the median RMSE for ML-SPGLM. As an additional measure, we provide \textit{RRMSE$_a$}, which is same as \textit{RRMSE$_m$} with the only difference being use of averages instead of medians across replicates. Similarly, the measures \textit{RL$_m$} and \textit{RL$_a$}, based on relative length of the uncertainty intervals, are reported. Although both types of result summaries are reported in the tables, for more robust conclusions, we discuss the comparison using median-based summaries only. Frequentist coverage probabilities (\textit{CP}) of posterior credible intervals and ML based confidence intervals are computed. These $95\%$ credible or confidence intervals (\textit{CI}) are constructed using symmetric quantiles based on posterior samples for Dir-SPGLM and Wald method for ML-SPGLM. 

\subsection{Results}
\label{sec:sim_results}

Table \customref{tab:beta-simulation-study} compares Dir-SPGLM estimates for the regression parameters $\beta_j$ to corresponding ML-SPGLM estimates in both small and large sample settings and under both simulation scenarios. As expected, both ML-SPGLM and Dir-SPGLM estimators have some bias in the small sample case, whereas both are approximately unbiased in the large sample regime. However, the Dir-SPGLM estimator is more efficient in all settings for the small sample case, with a $11\%$ to $21\%$ lower RMSE and $7\%$ to $12\%$ lower CI length. For large samples, both estimators have comparable efficiency with the Dir-SPGLM estimator being slightly more efficient. Frequentist coverage probabilities for the regression parameters based on both methods are close to the $0.95$ nominal level. Table \customref{tab:f0-small-sample-study} reports the same for $f_0(s_\ell), \ l= 0, \ldots, 5$, in small sample setting. For $f_0(0), \ldots,f_0(3)$, both methods report similar performance as for the regression parameters in terms of frequentist (un)biasedness and statistical efficiency. However, the Dir-SPGLM estimator performs favorably in estimating $f_0(4)$ and $f_0(5)$, underscoring the limitation of maximum likelihood estimation of $f_0(s_\ell)$ in sparse-data scenarios when only a limited number of observations are available for $y = s_\ell$. Also, ML-SPGLM does not naturally provide uncertainty quantification for $f_0(s_\ell)$ whereas the proposed Dir-SPGLM model provides the posterior credible intervals as a by-product of MCMC simulation. Overall, the proposed Dir-SPGLM model compares favorably with the ML-SPGLM model.

\begin{center}
\begin{table*}[!h]
\caption{Small and Large sample simulation results for regression parameters. \textit{Notes:} Est$_a$, RRMSE$_a$ and RL$_a$ (Est$_m$, RRMSE$_m$ and RL$_m$) are average (median) of Est(s), RRMSE(s) and RL(s) obtained from $R=2,\!000$ simulated data sets. Abbreviations: n -- Sample size; S -- Scenario as per Section \customref{section:5.1}; Par -- Parameter; Est -- Estimate; RRMSE -- Relative (w.r.t.\ ML-SPGLM) root mean square error; CP -- Coverage probability with $\alpha$ level $0.95$; RL -- Relative length (w.r.t.\ ML-SPGLM) for credible/confidence interval.}
\label{tab:beta-simulation-study}
\begin{tabular*}{\textwidth}{@{\extracolsep{\fill}}cccccccccccc}
\hline 
n & S & Par & Method & True & Est$_a$ & RRMSE$_a$ & RL$_a$ &Est$_m$ & RRMSE$_m$ & RL$_m$ & CP  \\
\hline
\multirow{8}{*}{25}  & \multirow{4}{*}{1} & $\beta_0$ & ML-SPGLM & -0.7 & -0.78 & 1.00 & 1.00 & -0.74 & 1.00 & 1.00  & 0.97\\
  &  & & Dir-SPGLM &  &-0.76 & 0.82 & 0.91 & -0.74 & 0.89 & 0.93 & 0.97\\
  \cline{3-12}
& & $\beta_1$ & ML-SPGLM & \ 0.2 & \ 0.20 & 1.00 & 1.00 & \ 0.19 & 1.00 & 1.00 & 0.93\\
&  & & Dir-SPGLM &  &\ 0.16 & 0.79 & 0.92 & \ 0.17 & 0.83 & 0.93 & 0.97\\
\cline{2-12}
 & \multirow{4}{*}{2} & $\beta_0$ & ML-SPGLM & -0.7 & -0.81 & 1.00 &1.00 & -0.77 & 1.00 & 1.00 & 0.97 \\
  &  & & Dir-SPGLM & & -0.77  & 0.75 & 0.87 & -0.75 & 0.85 & 0.90 & 0.97 \\
  \cline{3-12}
& & $\beta_1$ & ML-SPGLM & \ 0.2 & \ 0.18 & 1.00 & 1.00 & \ 0.18 & 1.00 & 1.00 & 0.94\\
&  & & Dir-SPGLM & & \ 0.14 & 0.73 & 0.86 & \ 0.14 & 0.79 & 0.88 &  0.97\\
\hline 
\multirow{8}{*}{250}  & \multirow{4}{*}{1} & $\beta_0$ & ML-SPGLM & -0.7 & -0.71 & 1.00 & 1.00 & -0.71 & 1.00 & 1.00 & 0.97\\
  &  & & Dir-SPGLM &  & -0.71 & 1.00 & 0.99 & -0.71 & 0.98 & 0.99 & 0.96\\
  \cline{3-12}
& & $\beta_1$ & ML-SPGLM & \ 0.2 & \ 0.20 & 1.00 & 1.00 & \ 0.20 & 1.00 & 1.00 & 0.96\\
&  & & Dir-SPGLM &  &  \ 0.20 & 0.99 & 0.99 & \ 0.20 & 1.00 & 0.98 & 0.96\\
\cline{2-12}
 & \multirow{4}{*}{2} & $\beta_0$ & ML-SPGLM & -0.7 & -0.71 & 1.00 & 1.00 & -0.71 & 1.00 & 1.00 & 0.97 \\
  &  & & Dir-SPGLM &  & -0.71 & 0.98 & 0.99 & -0.71 & 0.97 & 0.99 & 0.96 \\
  \cline{3-12}
& & $\beta_1$ & ML-SPGLM & \ 0.2 & \ 0.20 &1.00 &1.00 & \ 0.20 & 1.00 & 1.00 & 0.96 \\
&  & & Dir-SPGLM &  &\ 0.20 &0.97 & 0.98 & \ 0.20 & 0.96 & 0.98 & 0.96 \\
\hline
\end{tabular*}
\end{table*}
\end{center}

\begin{center}
\begin{table*}[!h]
\caption{Small sample simulation results for $f_0$. \textit{Notes:} Abbreviations are same as Table \customref{tab:beta-simulation-study}. N/A -- Not available. }
\label{tab:f0-small-sample-study}
\begin{tabular*}{\textwidth}{@{\extracolsep{\fill}}cccccccccc}
\hline
n & Scenario & Parm & Method & Truth & Est$_a$ & RRMSE$_a$ & Est$_m$ & RRMSE$_m$ & CP  \\
\hline 
\multirow{24}{*}{25}  & \multirow{12}{*}{1} & $f_0(0)$ & ML-SPGLM& 0.367 & 0.291 & 1.00 & 0.307 & 1.00 & N/A\\
  &  &  & Dir-SPGLM & & 0.364 & 0.59 & 0.368 & 0.71 & 0.94\\
\cline{3-10}
  & & $f_0(1)$ & ML-SPGLM & 0.368 & 0.454 & 1.00 & 0.414 & 1.00 & N/A \\
  &  &  & Dir-SPGLM & &0.386 &0.58 & 0.369 & 0.86 & 0.92\\
\cline{3-10}
  & & $f_0(2)$ & ML-SPGLM& 0.185& 0.220 & 1.00 & 0.236 & 1.00 & N/A \\
  &  &  & Dir-SPGLM & &0.168 & 0.62 & 0.159 & 0.60 & 0.95 \\
\cline{3-10}
  & & $f_0(3)$ & ML-SPGLM& 0.062&  0.034&1.00 & 0.000 & 1.00 & N/A \\
  &  &  & Dir-SPGLM & &0.058 & 0.52 & 0.051 & 0.40 & 0.92\\
\cline{3-10}
  & & $f_0(4)$ & ML-SPGLM& 0.015 & 0.001 & 1.00 & 0.000 & 1.00 & N/A\\
  &  &  & Dir-SPGLM & &0.018 & 1.00 & 0.014 & 0.34 & 0.97\\
\cline{3-10}
  & & $f_0(5)$ & ML-SPGLM& 0.003& 0.000 & 1.00 & 0.000 & 1.00 & N/A \\
  &  &  & Dir-SPGLM & &0.006 & 2.44 & 0.005 & 0.67 & 0.98\\
\cline{2-10}
  & \multirow{12}{*}{2} & $f_0(0)$ & ML-SPGLM & 0.471 & 0.397 & 1.00 & 0.409 & 1.00 & N/A \\
  &  &  & Dir-SPGLM & & 0.458 & 0.58 & 0.461 & 0.66 & 0.89 \\
\cline{3-10}
  & & $f_0(1)$ & ML-SPGLM & 0.232 & 0.288 & 1.00 & 0.259 & 1.00 & N/A \\
  &  &  & Dir-SPGLM & &0.246 &0.58 & 0.240 & 0.83 & 0.88 \\
\cline{3-10}
  & & $f_0(2)$ & ML-SPGLM & 0.172 & 0.236 & 1.00 & 0.240 & 1.00 & N/A \\
  &  &  & Dir-SPGLM & & 0.178 & 0.60 & 0.164 & 0.58 & 0.91 \\
\cline{3-10}
  & & $f_0(3)$ & ML-SPGLM & 0.085 & 0.070 & 1.00 & 0.062 & 1.00 & N/A \\
  &  &  & Dir-SPGLM & & 0.076 & 0.58 & 0.070 & 0.38 & 0.88\\
\cline{3-10}
  & & $f_0(4)$ & ML-SPGLM & 0.031 &0.007 & 1.00 & 0.000 & 1.00 & N/A \\
  &  &  & Dir-SPGLM & & 0.029 & 0.77 & 0.019 & 0.51 & 0.90\\
\cline{3-10}
  & & $f_0(5)$ & ML-SPGLM & 0.009 &0.000 &1.00 & 0.000 & 1.00 & N/A \\
  &  &  & Dir-SPGLM & & 0.010 & 1.44 & 0.006 & 0.46 & 0.93\\
\hline
\end{tabular*}
\end{table*}
\end{center}
Finally, we compare estimation of selected exceedance probabilities. For the ML-SPGLM, we estimate exceedance probabilities by plugging in maximum likelihood estimates:
\begin{equation}
\label{eqn:ex_est_ml}
   p(y \geq y_0 \mid x) \estimates p(y \geq y_0 \mid x; \beta^{(mle)}, f^{(mle)}_0) \ .  
\end{equation}
For Dir-SPGLM, we compute posterior predictive probabilities for the exceedance events defined as:
\begin{equation}
\label{eqn:ex_est_dir}
   p(y \geq y_0 \mid x) =  \int p(y \geq y_0 \mid x, \beta, f_0) \ p(\beta, f_0 \mid \mathcal{D}_n) d\beta d f_0 \estimates (1/B) \sum_{b = 1}^B p(y \geq y_0 \mid x, \beta^{(b)}, f^{(b)}_0) \ ,
\end{equation}
where $(\beta^{(b)}, f^{(b)}_0)$ denotes the $b$th posterior Monte Carlo sample. The comparison is performed for $y_0 = 2$ (moderate case) and $4$ (severe case) given different values of $x$. 
For the moderate case, both ML-SPGLM and Dir-SPGLM estimators have comparable performance in terms of frequentist (un)biasedness, with the Dir-SPGLM estimator being slightly more efficient. Performance of the ML-SPGLM estimator deteriorates as we move from moderate to severe exceedance events. In the case of severe exceedance events, the Dir-SPGLM compares favorably. The graphics are provided in Appendix B (Figures 5 and 6) of the Supplementary Information.

\subsection{Additional simulation studies}
We provide additional simulation results in the Supplementary Information. Figure 3 in Appendix B shows the Rao–Blackwellized marginal posterior density (\customcite{casella1996rao}) estimate with a $95\%$ credible band for the regression parameters. Figure 4 in Appendix B demonstrates consistency of the Dir-SPGLM estimates as a function of the sample size.  Table 1 in Appendix C reports large sample simulation results for $f_0$. 

\section{THE AHEAD STUDY}
\label{sec6} 
We fit both inference models for the AHEAD data set following the same setup as in the simulations reported in Section \customref{section:5.2}. The only change being the use of $10,\!000$ MCMC iterations for the Dir-SPGLM. We discard the first $3,\!000$ samples as initial transient, and use the remaining $B = 7,\!000$ posterior Monte Carlo samples for the reported inference.
\subsection{Predictive Inference}
A comparison of prediction under Dir-SPGLM versus ML-SPGLM  is implemented in two ways. First, we perform a small sample study in which both models are fitted based on a small training data set of size $n = 100$, randomly sampled from the complete AHEAD data of size $6,\!441$. We then assess prediction accuracy on the held-out test data set of size $6,\!341$. For comparison, we focus on estimating probabilities of exceedance events at moderate and severe difficulty in daily activities, i.e. $p(y \geq y_0 \mid x)$ with $y_0 = 2$ and $4$, respectively. We estimate these quantities using (\customref{eqn:ex_est_ml}) and (\customref{eqn:ex_est_dir}). The AUC (area under the curve) values for moderate exceedance are comparably similar: $0.70$ for ML-SPGLM and $0.71$ for Dir-SPGLM. In the case of severe exceedance, the AUC value is $0.75$ for ML-SPGLM and an improved $0.79$ for Dir-SPGLM. The ROC plot highlights that Dir-SPGLM consistently improves over ML-SPGLM. This performance difference is largest for the severe exceedance event. In fact, in sparse data cases, AUC under the ML-SPGLM can get close to $0.50$. The ROC plots are provided in Appendix D (Figures 7 and 8) of the Supplementary Information.

\begin{figure}[ht]
    \centering
     \includegraphics[height = 9cm, width = 12cm]{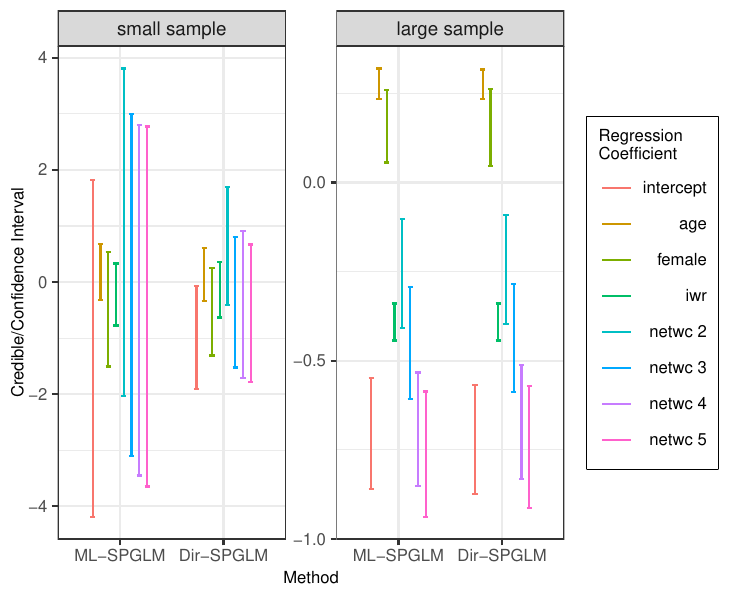}
    \caption{AHEAD study. Credible/Confidence intervals for regression parameters by methods based on small sample and large sample (full data set). \textit{iwr} (immediate word recall), and \textit{netwc j} (net worth category \textit{j}), \textit{j = 1, 2, \dots, 5}. 
    \label{fig3}}
\end{figure}

Second, we compare the precision in estimation using confidence and credible intervals for the regression coefficients. Figure \customref{fig3} shows these uncertainty intervals respectively for ML-SPGLM and Dir-SPGLM. In addition, Table \customref{tab:CI_diff_AHEAD} reports the relative interval lengths (w.r.t. ML-SPGLM) i.e., ratio of the interval length for Dir-SPGLM versus the ML-SPGLM. Notably, in the case of small training samples, the intervals are considerably narrower for Dir-SPGLM as compared to ML-SPGLM, with a reduction ranging from $6\%$ to $69\%$. This discrepancy reduces to a negligible difference with large training samples. This underscores that ML-SPGLM estimates can exhibit substantial uncertainty, particularly in scenarios involving small sample sizes, whereas Dir-SPGLM estimates offer narrower uncertainty intervals. In addition, we provide point estimates and uncertainty intervals for the model parameters in Appendix E (Tables 2 and 3) of the Supplementary Information.  

\begin{center}
\begin{table*}[!h]
\caption{AHEAD study. Small and large training sample performance using 95\% credible/confidence intervals of the regression parameters. The table shows the relative interval lengths for Dir-SPGLM (w.r.t.\ ML-SPGLM). \textit{Note:} See also Figure \customref{fig3}.}
\label{tab:CI_diff_AHEAD}
\begin{tabular*}{\textwidth}{@{\extracolsep{\fill}}ccccccccc}
\hline
    \multirow{2}{*}{Training sample size} & \multicolumn{8}{c}{Relative uncertainty interval lengths} \\
\cline{2-9} 
      & $\beta_0$ & $\beta_1$ & $\beta_2$ & $\beta_3$ & $\beta_4$ & $\beta_5$ & $\beta_6$ & $\beta_7$\\ 
  \hline
small & 0.31 & 0.94 & 0.76 & 0.89 & 0.36 & 0.38 & 0.42 & 0.38 \\ 
  large & 0.99 & 0.98 & 1.06 & 1.00 & 1.00 & 0.97 & 1.01 & 0.97 \\ 
   \hline
\end{tabular*} 
\end{table*}
\end{center}

\subsection{Uncertainty in exceedance probability estimates} 
\label{sec:exccedance}
\begin{figure}[ht]
    \centering
    \includegraphics[height = 10cm, width = 12cm]{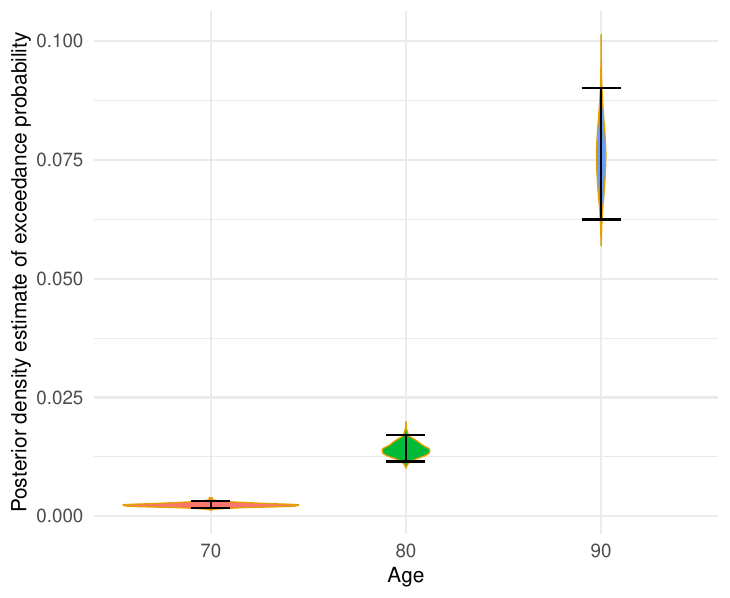}
    \caption{AHEAD study. Severe exceedance probabilities $\pi = p(y \geq 4 \mid x_{age}, \mathcal{D}_n)$ for $x_{age} = 70, 80$, and $90$. The vertically plotted density summarizes  uncertainty about the probability of the exceedance event as a function of 
    parameters $\phi = (\beta, f_0)$. That is, let $\pi_\phi = p(y \geq 4 \mid x_{age}, \phi)$ (with $\pi = \int \pi_\phi dp(\phi \mid \mathcal{D}_n)$). The violin plots show posterior distribution of $\pi_\phi$; a  peaked density indicates less posterior uncertainty about $\pi_\phi$.
    \label{fig4}}
\end{figure}

In decision-making problems, accurately quantifying uncertainty in exceedance probability estimates is an important consideration. However, inference under ML-SPGLM does not naturally report such uncertainties. In contrast, the Dir-SPGLM readily provides uncertainty quantification for any function of the model parameters as a byproduct of the generated posterior Monte Carlo samples. We focus here on both moderate and severe exceedance probabilities for different ages, i.e. $p(y \geq y_0 \mid x_{age}), \ y_0 = 2, 4, \mbox{ and } x_{age} = 70, 80, 90$. For short, we write model parameters as $\phi = (\beta, f_0)$. The $b$th posterior Monte Carlo sample for $p(y \geq y_0 \mid x_{age}, \phi)$ is obtained as
\[
   p\left(y \geq y_0 \mid x_{age} = t, \phi^{(b)}\right) = \frac{\underset{x: x_{age} = t}{\sum} p\left(y \geq y_0 \mid x, \phi^{(b)}\right)}{\sum_{x}1_{\{x: x_{age} = a\}}(x)} \ ,
\]
where $\phi^{(b)} = \left(\beta^{(b)}, f^{(b)}_0\right)$ denotes the $b$th sample from the posterior. Note that in this case, the design with respect to $x$ is considered to be fixed. Figure \customref{fig4} shows a kernel  density estimate
of the implied posterior distribution on severe exceedance probabilities, $p(y \geq 4 \mid x_{age}, \phi)$. A  peaked density indicates less posterior uncertainty about $p(y \geq 4 \mid x_{age}, \phi)$. As expected, with an increase in age, exceedance probability estimates increase non-linearly but with a decrease in precision. This pattern is comparatively less prominent in case of moderate exceedance (see Figure 9 of the Supplementary Information).

\section{DISCUSSION}
\label{sec7}
We have proposed a flexible Bayesian model-based approach to inference under a recently developed semi-parametric extension of the classical generalised linear model (SPGLM). The major contributions of the developed Dir-SPGLM are, first, it allows us to estimate and provide critically needed uncertainty quantification for the centering distribution and model-derived quantities such as exceedance probabilities, which may be important in clinical diagnostic or other decision-making settings. Second, it shows promise in small sample and sparse-data scenarios over the existing maximum likelihood-based methods. Third, it provides a launching pad for important future work; see last paragraph.

Our simulation study shows that the proposed Dir-SPGLM compares favourably with maximum likelihood inference (ML-SPGLM) in terms of (un)biased parameter estimates, coverage probabilities, and statistical efficiency in both small and large sample regimes, with considerable benefits in small sample settings. We have also discussed an application of the proposed method for inference in the AHEAD study. Results show that Dir-SPGLM improves predictive performance, in case of exceedance events, over ML-SPGLM, and this performance difference is largest for sparse-data scenarios. Credible intervals for model parameters are more precise under the Dir-SPGLM, again, especially in small sample settings. In addition, under the Dir-SPGLM, posterior inference for moderate and severe exceedance probabilities show that with increasing age, exceedance probabilities increase non-linearly but with a decrease in precision for the estimate. This pattern is most prominent for severe exceedance, as expected, since data are sparser for older ages and for higher exceedance thresholds.

For future work, building upon this current methodology, there are multiple research directions one could explore. One is a Bayesian inference approach for multi-variate or repeated measures SPGLM that will include random effects (or other latent variables); a Bayesian approach is especially attractive for this application because, under the ML paradigm, similar extensions are likely to be prohibitively difficult to implement using adaptive quadrature, the default computational approach for non-linear random effects regression models.  Second, the proposed Dir-SPGLM will serve as a foundation for a semi-parametric (continuous response) setting. Third, the SPGLM, married with our Bayesian approach to inference, unlocks opportunities to balance cost-effectiveness and precision in clinical study designs for finding specific population subgroups significantly affected by a particular condition, disease, or treatment response in cases where the response distribution follows a form not well captured by existing parametric models. This could be achieved by finding the minimum sample size required \customcitefr{riley2019minimumI}{riley2019minimumII}{archer2021minimum}{riley2021minimum}, in setting up clinical designs, while maintaining a permitted error level for our model-derived quantity of interest.



\section*{Acknowledgments}
This research was supported by the National Institutes of Health Grant 2 R01 HL094786.



\section*{Conflict of interest}

The authors declare no potential conflict of interests.

\section*{Data Availability Statement}
The data that support the findings in this paper are available in the University of Michigan Health and Retirement Study (\customcite{hrs}) repository at \href{https://hrsdata.isr.umich.edu/data-products/1993-ahead-core}{\textcolor{blue}{https://hrsdata.isr.umich.edu/data-products/1993-ahead-core}}. For accessing the curated data, kindly direct request to P. J. Rathouz (\href{mailto:paul.rathouz@austin.utexas.edu}{\textcolor{blue}{paul.rathouz@austin.utexas.edu}}).

\bibliography{Dir-SPGLM_Main_Manuscript}

\begin{thebibliography}{}

\bibitem[\protect\citeauthoryear{Archer, Snell, Ensor, Hudda, Collins, and
  Riley}{Archer et~al.}{2021}]{archer2021minimum}
Archer, L., K.~I. Snell, J.~Ensor, M.~T. Hudda, G.~S. Collins, and R.~D. Riley
  (2021).
\newblock {Minimum sample size for external validation of a clinical prediction
  model with a continuous outcome}.
\newblock {\em Statistics in Medicine\/}~{\em 40\/}(1), 133--146.

\bibitem[\protect\citeauthoryear{Casella and Robert}{Casella and
  Robert}{1996}]{casella1996rao}
Casella, G. and C.~P. Robert (1996).
\newblock {Rao-Blackwellisation of sampling schemes}.
\newblock {\em Biometrika\/}~{\em 83\/}(1), 81--94.

\bibitem[\protect\citeauthoryear{Ferguson}{Ferguson}{1973}]{ferguson1973bayesian}
Ferguson, T.~S. (1973).
\newblock {A {B}ayesian analysis of some nonparametric problems}.
\newblock {\em The Annals of Statistics\/}, 209--230.

\bibitem[\protect\citeauthoryear{Gelfand and Sahu}{Gelfand and
  Sahu}{1999}]{gelfand1999identifiability}
Gelfand, A.~E. and S.~K. Sahu (1999).
\newblock {Identifiability, improper priors, and Gibbs sampling for generalized
  linear models}.
\newblock {\em Journal of the American Statistical Association\/}~{\em
  94\/}(445), 247--253.

\bibitem[\protect\citeauthoryear{Ghosal and Van~der Vaart}{Ghosal and Van~der
  Vaart}{2017}]{ghosal2017fundamentals}
Ghosal, S. and A.~Van~der Vaart (2017).
\newblock {\em {Fundamentals of nonparametric {B}ayesian inference}},
  Volume~44.
\newblock Cambridge University Press.

\bibitem[\protect\citeauthoryear{Green and Richardson}{Green and
  Richardson}{2001}]{GreenRichardson01}
Green, P.~J. and S.~Richardson (2001).
\newblock {Modelling Heterogeneity with and without the {D}irichlet Process}.
\newblock {\em Scandinavian Journal of Statistics\/}~{\em 28\/}(2), 355--375.

\bibitem[\protect\citeauthoryear{HRS}{HRS}{1993}]{hrs}
HRS (1993).
\newblock {Health and Retirement Study, (AHEAD 1993 Core) Public Use Dataset.
  Produced and distributed by the University of Michigan with funding from the
  National Institute on Aging (grant number NIA U01AG009740). Ann Arbor, MI}.

\bibitem[\protect\citeauthoryear{Huang}{Huang}{2014}]{huang2014joint}
Huang, A. (2014).
\newblock {Joint estimation of the mean and error distribution in generalized
  linear models}.
\newblock {\em Journal of the American Statistical Association\/}~{\em
  109\/}(505), 186--196.

\bibitem[\protect\citeauthoryear{Kossin, Knapp, Olander, and Velden}{Kossin
  et~al.}{2020}]{kossin2020global}
Kossin, J.~P., K.~R. Knapp, T.~L. Olander, and C.~S. Velden (2020).
\newblock {Global increase in major tropical cyclone exceedance probability
  over the past four decades}.
\newblock {\em Proceedings of the National Academy of Sciences\/}~{\em
  117\/}(22), 11975--11980.

\bibitem[\protect\citeauthoryear{Lijoi and Pr{\"u}nster}{Lijoi and
  Pr{\"u}nster}{2010}]{ghosal_2010}
Lijoi, A. and I.~Pr{\"u}nster (2010).
\newblock {The {D}irichlet process, related priors and posterior asymptotics}.
\newblock In N.~L. Hjort, C.~Holmes, P.~M{\"u}ller, and S.~G. Walker (Eds.),
  {\em {B}ayesian Nonparametrics}, pp.\  35–79. Cambridge University Press.

\bibitem[\protect\citeauthoryear{MacEachern}{MacEachern}{2000}]{maceachern;2000}
MacEachern, S.~N. (2000).
\newblock {Dependent {D}irichlet processes}.
\newblock Technical report, Department of Statistics, The Ohio State
  University.

\bibitem[\protect\citeauthoryear{Maronge, Tao, Schildcrout, and
  Rathouz}{Maronge et~al.}{2023}]{maronge2023generalized}
Maronge, J.~M., R.~Tao, J.~S. Schildcrout, and P.~J. Rathouz (2023).
\newblock {Generalized case-control sampling under generalized linear models}.
\newblock {\em Biometrics\/}~{\em 79\/}(1), 332--343.

\bibitem[\protect\citeauthoryear{McCullagh and Nelder}{McCullagh and
  Nelder}{1989}]{mccullagh1989generalized}
McCullagh, P. and J.~A. Nelder (1989).
\newblock {Generalized linear models 2nd edition chapman and hall}.
\newblock {\em London, UK\/}.

\bibitem[\protect\citeauthoryear{Paul, Adeyemi, Ghosh, Pokhrel, and Arif}{Paul
  et~al.}{2021}]{paul2021dynamics}
Paul, R., O.~Adeyemi, S.~Ghosh, K.~Pokhrel, and A.~A. Arif (2021).
\newblock {Dynamics of Covid-19 mortality and social determinants of health: a
  spatiotemporal analysis of exceedance probabilities}.
\newblock {\em Annals of Epidemiology\/}~{\em 62}, 51--58.

\bibitem[\protect\citeauthoryear{Quintana, M{\"u}ller, Jara, and
  MacEachern}{Quintana et~al.}{2022}]{quintana2022dependent}
Quintana, F.~A., P.~M{\"u}ller, A.~Jara, and S.~N. MacEachern (2022).
\newblock {The dependent {D}irichlet process and related models}.
\newblock {\em Statistical Science\/}~{\em 37\/}(1), 24--41.

\bibitem[\protect\citeauthoryear{Rathouz and Gao}{Rathouz and
  Gao}{2009}]{rathouz2009generalized}
Rathouz, P.~J. and L.~Gao (2009).
\newblock {Generalized linear models with unspecified reference distribution}.
\newblock {\em Biostatistics\/}~{\em 10\/}(2), 205--218.

\bibitem[\protect\citeauthoryear{Riley, Debray, Collins, Archer, Ensor, van
  Smeden, and Snell}{Riley et~al.}{2021}]{riley2021minimum}
Riley, R.~D., T.~P. Debray, G.~S. Collins, L.~Archer, J.~Ensor, M.~van Smeden,
  and K.~I. Snell (2021).
\newblock {Minimum sample size for external validation of a clinical prediction
  model with a binary outcome}.
\newblock {\em Statistics in medicine\/}~{\em 40\/}(19), 4230--4251.

\bibitem[\protect\citeauthoryear{Riley, Snell, Ensor, Burke, Harrell~Jr, Moons,
  and Collins}{Riley et~al.}{2019a}]{riley2019minimumI}
Riley, R.~D., K.~I. Snell, J.~Ensor, D.~L. Burke, F.~E. Harrell~Jr, K.~G.
  Moons, and G.~S. Collins (2019a).
\newblock Minimum sample size for developing a multivariable prediction model:
  Part i--continuous outcomes.
\newblock {\em Statistics in medicine\/}~{\em 38\/}(7), 1262--1275.

\bibitem[\protect\citeauthoryear{Riley, Snell, Ensor, Burke, Harrell~Jr, Moons,
  and Collins}{Riley et~al.}{2019b}]{riley2019minimumII}
Riley, R.~D., K.~I. Snell, J.~Ensor, D.~L. Burke, F.~E. Harrell~Jr, K.~G.
  Moons, and G.~S. Collins (2019b).
\newblock {Minimum sample size for developing a multivariable prediction model:
  PART II-binary and time-to-event outcomes}.
\newblock {\em Statistics in medicine\/}~{\em 38\/}(7), 1276--1296.

\bibitem[\protect\citeauthoryear{Soldo, Hurd, Rodgers, Wallace, et~al.}{Soldo
  et~al.}{1997}]{soldo1997asset}
Soldo, B.~J., M.~D. Hurd, W.~L. Rodgers, R.~B. Wallace, et~al. (1997).
\newblock {Asset and health dynamics among the oldest old: An overview of the
  AHEAD study}.
\newblock {\em Journals of Gerontology Series B\/}~{\em 52}, 1--20.

\bibitem[\protect\citeauthoryear{Tajima, Tanaka, and Wu}{Tajima
  et~al.}{2021}]{tajima2021empirical}
Tajima, R., H.~Tanaka, and C.~Wu (2021).
\newblock {An Empirical Method for Estimating Source Vicinity Ground-Motion
  Levels on Hard Bedrock and Annual Exceedance Probabilities for Inland Crustal
  Earthquakes with Sources Difficult to Identify in Advance}.
\newblock {\em Bulletin of the Seismological Society of America\/}~{\em
  111\/}(5), 2408--2425.

\bibitem[\protect\citeauthoryear{Taylor and Yu}{Taylor and
  Yu}{2016}]{taylor2016using}
Taylor, J.~W. and K.~Yu (2016).
\newblock {Using auto-regressive logit models to forecast the exceedance
  probability for financial risk management}.
\newblock {\em Journal of the Royal Statistical Society Series A: Statistics in
  Society\/}~{\em 179\/}(4), 1069--1092.

\bibitem[\protect\citeauthoryear{Wurm and Rathouz}{Wurm and
  Rathouz}{2018}]{wurm2018semiparametric}
Wurm, M.~J. and P.~J. Rathouz (2018).
\newblock {Semiparametric generalized linear models with the gldrm package}.
\newblock {\em The R journal\/}~{\em 10\/}(1), 288.

\end{thebibliography}

\end{document}